\documentstyle[12pt]{article}

\newcommand\pubnumber{TIFR-TH-99-41}
\newcommand\pubdate{August 1999}

\def\Title#1{\begin{center} {\Large #1 } \end{center}}
\def\Author#1{\begin{center}{ \sc #1} \end{center}}
\def\Address#1{\begin{center}{ \it #1} \end{center}}

\def\TIFR{Department of Theoretical Physics\\
    Tata Institute of Fundamental Research\\
    Homi Bhabha Road, Mumbai, India, 400005}
\newcommand\pubblock{\rightline{\begin{tabular}{l} hep-th/9908077\\
\pubnumber\\
         \pubdate  \end{tabular}}}
\newenvironment{Abstract}{\begin{quotation} \begin{center}
                       ABSTRACT
     \end{center}\bigskip  }{\end{quotation}}

\def\ket#1{|#1\rangle}
\def\cA{{\cal A}}
\def\cZ{{Z}}

\def\bZ{{\bf Z}}

\newcommand{\nc}{\newcommand}
\nc{\beq}{\begin{equation}} \nc{\eeq}{\end{equation}}
\nc{\beqa}{\begin{eqnarray}} \nc{\eeqa}{\end{eqnarray}}

\font\titlefont=cmcsc10 scaled \magstep3
\font\secfont=cmcsc10 scaled \magstep2
\font\subsecfont=cmr10 scaled \magstep2
\begin{document}
\begin{titlepage}
\pubblock

\vfill
\Title{\titlefont
Discrete Flux as  Quantum Hair}
\vskip0.5truein
\Author{Atish Dabholkar and Sandip P. Trivedi}
\Address{\TIFR}
\vskip0.5truein
\begin{Abstract}
We investigate  Yang-Mills theory  on a spatial torus at finite temperature  
in the presence of discrete electric and magnetic fluxes using the AdS/CFT 
correspondence. We  calculate the leading dependence of the  partition 
function on the  fluxes using the dual supergravity theory and comment upon 
the interpretation of these fluxes as discrete quantum hair for black holes in 
AdS spacetime.

\end{Abstract}
\medskip

\vfill
\end{titlepage}
\def\thefootnote{\fnsymbol{footnote}}
\setcounter{footnote}{0}
\section{\secfont Introduction}
In this paper we calculate the finite temperature 
partition function of  the Yang-Mills  theory with
sixteen supercharges on a torus 
as a function of discrete electric and magnetic fluxes. 
We perform this calculation  in the large $N$ limit using  the dual supergravity description in four and five dimensions \cite{Mald,AGMOO}.
In this description,  the thermal state of the gauge theory maps 
to a black hole \cite{Witt1} 
and we will show that the discrete flux in the gauge theory 
maps to a kind of quantum hair for the black hole. Thus, our calculation will 
reduce to determining how the partition function of a black hole depends on its
quantum hair.  

The study of discrete fluxes in nonabelian gauge theories was initiated
by 't Hooft \cite{'tHoo} following the analogies between confinement
and superconductivity. 
Magnetic lines of force are expelled from 
the superconducting ground state 
and  can penetrate the medium  only in the 
form of thin flux tubes. Magnetic flux tubes
exist as Nielsen-Olsen vortices and are locally stable.
If a pair of magnetic monopole and antimonopole is introduced in
the medium then the total flux between them is localized in
a flux tube that connects the two charges giving rise to a linear confining
potential between the two. 
Qualitatively, confinement  in nonabelian gauge theories
is  the electric analog of superconductivity
where the  electric flux tubes are locally stable and
can then be interpreted as QCD strings. 
To put this idea on a more quantitative basis, 't Hooft  considered
the gauge theory on a torus. This allows for the introduction of
topological discrete fluxes in the pure gauge theory without quarks
which are the analogues of electric and magnetic fluxes in the U(1) theory.
By examining the dependence of  the free energy on  these fluxes, 
one can  investigate  the stability of the flux tubes to gain 
insight about the  phases of the theory.  In this paper we investigate some
of these questions using the correspondence with supergravity
which provides an explicit realization of  these ideas.

By now a fair amount is known about the finite temperature SYM theory we will 
consider here. For example, the spatial  Wilson loop 
has been calculated in the theory and shows area law behavior thereby indicating  that the theory 
has locally stable flux tubes \cite{Mald1,ReYe}. 
In this paper we will follow the general framework
discussed in \cite{AhWi, Witt2} for including discrete flux in 
supergravity.
Our calculation will be done in an approximation
where the string tension of these flux tubes is much bigger than the size of the 
torus. As we will see, in this approximation, the dominant configurations contributing to the 
partition function have a simple interpretation in the gauge theory: 
they correspond to the worldsheets of flux tubes wrapping appropriate
two cycles of the torus. We will do the calculation in both the four-dimensional 
and the  five-dimensional theory. The latter,  at energies low compared to the 
temperature, should flow to the four-dimensional pure glue gauge theory 
 without supersymmetry. Unfortunately, as is well known, in the supergravity 
regime
we use here, there are extra Kaluza-Klein states in the theory which do not 
decouple \cite{Witt1, GrOo}.  
Thus, from the point of view of the non-supersymmetric gauge theory,
our calculation is at best an approximation, somewhat analogous to the 
strong coupling 
expansion in lattice gauge theory. 
In fact, as we will show, the dual supergravity 
yields an answer which, in its important quantitative features, 
 agrees with the strong coupling expansion. 
These features are  in accord with the expectation about confining
theories and   are 
likely to be universal.

It is also useful to consider the discussion in this paper from  the perspective of the supergravity theory. As was mentioned above, a black hole in the bulk is
 represented as a thermal 
state in the boundary gauge theory. The classical no-hair theorems of black hole
 physics then turn into a familiar statement in the gauge theory:  a thermal state is only 
characterized by its mass,  angular momentum,  and various conserved global charges. 
But as is well known
\cite{BGHHS, CPW}, black holes can also possess quantum hair. One might wonder what their
description is  in the boundary theory.
For the quantum  hair considered in this paper they turn out to be related to the
the discrete electric and magnetic fluxes of the gauge theory. 

This paper is organized as follows. The 
four-dimensional theory  with $N=4$ supersymmetric
is discussed in \S2. The AdS/CFT correspondence for the
blackhole geometry is reviewed in \S2.1,
discrete fluxes from the gauge theory  and supergravity points of view 
are described in \S2.2 and \S2.3 respectively, and   the calculation of the 
partition function is discussed in \S2.4. 
Nonsupersymmetric 
gauge theories in three and four dimensions are discussed in \S3 and \S4 respectively
starting with a  supersymmetric theory in one extra dimension
compactified on a circle with thermal boundary conditions that break
supersymmetry.

\section{{ \secfont Discrete Fluxes in the AdS/CFT Correspondence}}

\subsection{\subsecfont The Black Hole Geometry}
In this section  we consider the  $N=4$
supersymmetric $SU(N)$  Yang-Mills theory 
on a spatial torus ${\bf T^3}$ at  finite temperature
$T$.
The boundary spacetime $\bf M$ in this case has
topology ${\bf T^3} \times {\bf S^1}$. By the AdS/CFT correspondence
\cite{Mald, Witt1},
this boundary theory is dual to a bulk supergravity theory on 
${\bf B} \times {\bf S^5}$ where $\bf B$ is an Einstein manifold
that has $\bf M$ as the boundary at infinity.
The relevant manifold $\bf B$ is well studied and 
 corresponds to a Euclidean Schwarzschild black hole in AdS space\cite{HaPa}.
In Poincare coordinates, the 
metric is\footnote{Since we are working in a situation where the spatial boundary  is  ${\bf T}^3$ as opposed to ${\bf S}^3$, the relevant solution is
obtained by taking the large mass limit of the metric in \cite{Witt1}.}

\beq
\label{solbh}
ds^2=({r^2 \over R^2}-{c G_5 M \over r^2} ) dx_0^2 + 
{dr^2 \over ({r^2 \over R^2}
-{c G_5 M \over r^2})} + {r^2 \over R^2} (dx_1^2 +dx_2^2+dx_3^2),
\eeq
where $R=(4\pi g_s \alpha' N)^{1/4}$ is the radius of curvature of AdS space, 
$G_5=G_{10}/ (R^5 \pi^3)$ is the five-dimensional  Newton's constant,  
$M$ is the mass of the black hole, and $c={ 32 \over 3 \pi}$. 
The black hole horizon is at
\beq
\label{hor}
r_H= ( c~ G_5 M~ R^2 )^{1/4}
\eeq
and the temperature is 
\beq
\label{temp}
T ={1 \over \pi}~ r_H/R^2.
\eeq

The coordinates ${x_i}$ ($i=1, 2, 3$) parametrize a 3-torus.
For simplicity we consider a cubic torus
of size $L$
so that  $x_i$
are identified with $x_i + L$.
Note that
the physical size of the torus at the radial position $r$ in the bulk is
$rL/R$.  In the black hole geometry,
$r$ is
bounded from below by the radius of black hole horizon $r_H$.
As long as $r_H L/R$ is large compared to the string scale, we  expect
that the dual supergravity description will be well-defined.
By contrast, at zero temperature  ($M=0$ in the above formulae),
 the size of the torus becomes vanishingly small as $r$ goes to zero.
The supergravity description then becomes inadequate at small $ r$
because the
string modes that wind around the torus become massless near $r=0$
and have to be included  in the low energy description.

\subsection{\subsecfont Fluxes in the Boundary Gauge Theory}

The finite temperature partition function of the Yang-Mills  theory
on the boundary
is obtained by computing the Euclidean path integral on the 4-torus
$\bf T^3 \times \bf S^1$. In the $N=4$ supersymmetric theory, the center 
${\bf Z}_N$ of $SU(N)$ 
acts trivially because all fields transform 
in the adjoint representation.  Therefore, one is  really dealing with an
$SU(N)/{\bf Z}_N$ gauge bundle on a 4-torus. Such
gauge bundles are labeled by six topological invariants
$ n_{\mu\nu}=-n_{\nu\mu}$ ($\mu, \nu =0,\ldots, 3$)
which are all integers modulo $ N$.
We would like to calculate the dependence of the partition function 
on these  integers.

Let us briefly review the origin of these integers and their
physical interpretation \cite{'tHoo, Pres}. 
In the path integral, one sums over
field configurations that are  periodic
up to a gauge transformation\footnote{ The  fermions are actually periodic
on $\bf T^3$ and  antiperiodic on  $\bf S^1$ but we will ignore this
distinction in this subsection}.  This allows for twisted gauge fields.
 For example, in the $x_1-x_2$ plane with
$ x_0$ and $ x_3$ fixed, a field $ \Phi$ satisfies the boundary condition
\beqa
\label{bc}
\Phi(L, x_2) &=& \Omega_1(x_2) \Phi(0, x_2) \nonumber\\
\Phi(x_1, L) &=& \Omega_2(x_1) \Phi(x_1, 0).
\eeqa
Here   $\Omega\Phi$
denotes schematically  the appropriate 
gauge transformation of $ \Phi$ by $ \Omega$:
$ A_{\mu} \rightarrow
\Omega A_{\mu} \Omega^{-1} - i\partial_{\mu}\Omega \Omega^{-1} $
for the gauge fields,
$ \lambda \rightarrow \Omega \lambda\Omega^{-1}$
for the gauginoes,
and similarly for the scalars.
For consistency, the gauge transformations must satisfy the  cocycle
 condition 
\beqa
\label{consistency}
\Phi(L, L) &=& \Omega_1(L) \Phi(0, L) = \Omega_1(L) \Omega_2(0)
\Phi(0, 0) \nonumber\\
 &=& \Omega_2(L) \Phi(L, 0) = \Omega_2(L) \Omega_1(0) \Phi(0, 0) .
\eeqa
Therefore, $ \Omega_1(L) \Omega_2(0)$ and
$ \Omega_2(L) \Omega_1(0)$ must be  equal up to
an element of the center ${\bf Z}_N$,
\beq
\label{cocycle}
\Omega_1(L) \Omega_2(0)
= \exp{({2\pi i n_{12} \over N}) } \Omega_2(L) \Omega_1(0).
\eeq
The integer $ n_{12}$ is defined modulo $ N $ and is
a topological invariant because it cannot be changed by
a  periodic gauge transformation or by smooth deformations of the field.
There are  altogether six independent  integers $ n_{\mu\nu}$
corresponding to the six 
2-cycles of the 4-torus
 that completely specify the topological class 
of the gauge fields. 

The integers $ \{n_{ij}\}$ ($i, j =1, 2, 3$) are related to the magnetic flux on the torus.
For example, consider  the contractible Wilson loop
$\Omega_1(L) \Omega_2(0) \Omega^{-1}_1(0)
\Omega^{-1}_2(L)$  that measures
the total magnetic flux passing through the $x_1-x_2$ plane.
We immediately see from the cocycle condition eq.(\ref{cocycle}) that  
there is $ n_{12}$ units of 
magnetic flux in the $x_3$ direction  for the twisted gauge fields
considered  above.
In general, one can define
the integer $ m_i \equiv {1\over 2} \epsilon_{ijk} n_{jk}$
as the  magnetic  flux in the $i$th direction.

The physical interpretation  of $n_{0i}$ is clearest
in the Hamiltonian formalism. In the gauge
$A_0=0$, the
theory has a residual invariance under time-independent gauge
transformations.
States in the
physical Hilbert space must furnish a representation of this
invariance group.
For a gauge transformation $\Omega (\bf x )$
that is  continuously connected to the identity, one must choose
the trivial representation for the corresponding operator
consistent with Gauss Law:
\beq
\label{trivial}
{\hat \Omega}(\bf x) \ket{\psi} = \ket{\psi}
\eeq
However, for the $ SU(N)/{\bf Z}_N$ theory on  $ \bf T^3$
there are quasiperiodic 
gauge transformations that are is not continuously connected to the identity. 
A quasiperiodic gauge transformation is
periodic  modulo
an element  of the center.  For example, consider
a gauge transformation that is  quasiperiodic in the $ x_3$ direction
\beq
\label{quasi}
\Omega[{k_3}](x_1, x_2, L) = 
\Omega[k_3] (x_1, x_2, 0) \exp{({2\pi i k_3 \over N}) }
\eeq
and periodic in the $ x_1, x_2$ directions.
Two such gauge transformations that are labeled by the same integer
$ k_3$ have the same action on physical states because  they
differ by a homotopically trivial gauge transformation that leaves
such states invariant as in eq.({\ref{trivial}}).  
Such gauge transformations obviously generate a ${\bf Z}_N$ group
and moreover commute
with the Hamiltonian.
One can choose the physical states to be eigenstates of this
gauge transformation
\beq
\label{rep}
{\hat \Omega}[k_3] \ket{\psi} = 
\exp{ ({2\pi i k_3 e_3 \over N})}\ket{\psi},
\eeq
for some integer $ e_3$ modulo $ N$.
To see that the integer $ e_3$ can be interpreted as the electric flux in the 
$ x_3$ direction,
consider  the action of $ \Omega[k_3]$ on a Wilson loop
\beq
\label{wilson}
{\bf \cal A}(C_3) = {\rm tr }P \exp (i \int_{C_3} dx^{\mu} A_{\mu})
\eeq
that runs along a curve $C_3$  in the $ x_3$ direction
at $ x_1= x_2 =0$.
It is clear
that 
\beq
\label{comm}
 {\hat \Omega}[k_3] \cA(C_3)= \cA(C_3) {\hat \Omega}[k_3] 
\exp{({2\pi i k_3\over N})}.
\eeq
Thus ${\cal A} (C_3)$ acting on a state increases the value
of $ e_3$ by one unit. 
Since  the Wilson loop creates a line of electric flux,
it is natural to regard $ e_3$ as the discrete electric flux. 

One can project onto states with a well defined electric
flux $ e_3$ by using
the projection operator
\beq
\label{projector}
{\cal P} (e_3) = {1\over N} \sum_{k_3} \exp{ ({-2\pi i k_3 e_3 \over N})}
{\hat \Omega}[k_3].
\eeq
The finite temperature partition sum over states with a
specified value of $ e_3$ is then given by
\beq
\label{trace}
\cZ(e_3) \equiv {\rm Tr} [\,{\cal P}\left( e_3 \right) \exp{( -H/T)}]=
{1\over N} \sum_{k_3} \exp{ ({-2\pi i k_3 e_3 \over N})}\,
{\rm Tr}\,[{\hat \Omega}[k_3] \exp{( -H/T)}].
\eeq
It is easy to see that each term 
${\rm Tr} [\,{\hat \Omega}[k_3] \exp{ (-H/T)}]$ in the sum can be expressed
as a Euclidean functional integral   over gauge
fields that are twisted in the $ x_0-x_3$ plane 
as in eq.({\ref{bc}})with $ n_{03}=k_3$. The partition function
$ \cZ(e_3)$
is then obtained as a discrete Fourier transform using
eq.(\ref{trace}). Thus, the integer $ k_3$ is conjugate to 
the electric flux $ e_3$.

More generally, the electric fluxes $ \{e_i\}$  are
 the quantum numbers of the ${ \bf Z}^3_N$ electric
symmetry and  are conjugate to the integers $\{k_i \equiv n_{0i}\}$.  
In addition, we also have $ {\bf Z}^3_N$ magnetic symmetry.
The integers $ m_i$ that we defined earlier
can equivalently be regarded as the quantum numbers
of the this magnetic symmetry in the dual description. 

\subsection{\subsecfont Fluxes in the Bulk Supergravity}

We would like to identify
the topological $ \bZ^6_N$ symmetry
and the  quantum numbers that correspond to
the discrete electric and magnetic fluxes  in 
the dual  supergravity  theory. 
We will follow the general framework
described in \cite{Witt2, AhWi}  although the specific context here
is somewhat different.

Consider, for example, the electric flux $ e_3$ which is  the quantum number
of the $ {\bZ}_N$ symmetry that is generated by  ${\hat \Omega}[k_3] $.
A defining property  of $ {\hat \Omega}[k_3] $ is the commutation
relation eq.(\ref{comm}) with the Wilson loop
$ {\cal A} (C_3)$. 
In the AdS/CFT correspondence, a Wilson loop $ \cA (C)$ that runs
along a curve $ C$ in the Yang-Mills theory is identified with the boundary
of a fundamental string worldsheet  $ D$ that extends in the bulk geometry
\cite{Mald1, ReYe}.
For ${\cal A} (C_3)$, the corresponding
string worldsheet $ D_{3r}$
extends,
at a fixed Euclidean time $ x_0$, in the $ x_3-r$ plane
such that $ \partial D_{3r} = C_3$. The worldsheet couples to
the NS-NS 2-form field $ B^{NS}$ through the coupling
\beq
\exp{(i\int_{D_{3r}} B^{NS})}
\eeq 
In particular, if we have a nonzero expectation value 
\beq
a = \int_{D_{3r}} B^{NS}
\eeq
in the bulk then the worldsheet would pick up a phase $ e^{ia}$. 
It follows from the commutation relations eq.(\ref{comm}) that
the operator that corresponds to ${\hat \Omega}[k_3] $
would shift this expectation value
\beq
\label{shift}
a\rightarrow a + 2\pi k_3/ N.
\eeq
We will see that in the supergravity theory,
 there is a natural candidate for this $ \bZ_N$ shift
symmetry. Obviously, the generator of this symmetry  would be 
the momentum conjugate to   $a$ which we can find from the
classical action and the corresponding translation operator can then
be identified with
${\hat \Omega}[k_3] $.

The relevant part of the supergravity Lagrangian on the 
 Euclidean AdS space
is
\beq
\label{lagra}
{\cal L} = {1\over 2 g_s^2} \left( |dB^{NS}|^2 + |dB^{R}|^2  \right) +  
{i N\over 4\pi}\left( B^{NS} \wedge dB^{R}  -B^{R} \wedge dB^{NS} \right).
\eeq
The first two terms are the usual kinetic terms restricted to the AdS part.
The last two terms are topological and 
arise from  the Chern-Simons coupling in ten dimensions of the two B-fields
to the 5-form field strength $ G_5$. 
The factor of $ N$ arises from the integration of 
 $ G_5$ over the sphere ${\bf S}^5$ and
corresponds to the total
5-form flux of  $ N$ D3-branes.

Since we are interested in spatially constant modes of the $ B$ fields,
we can integrate over space for the relevant modes and reduce this problem to
particle mechanics.
Let us define 
\beq
b = \int_{\Sigma_{12}} B^{R},
\eeq
for a nontrivial cycle $ \Sigma_{12}$ along $ x_1-x_2$ directions.
Then the  Lagrangian after an integratiioin by parts reduces to
\beq
\label{lagra2}
{L} = {1 \over 2 } \left( \mu_1 {\dot a}^2 + \mu_2 {\dot b}^2 \right)  +  
{i N\over 2\pi} \,{{a}{\dot b}} \, .
\eeq
Here the
dot indicates a derivative with respect to the Euclidean time.
The regularized effective masses $\mu_1, \mu_2$ arise from integrating
the $ r$ dependence of the metric in the original Lagrangian
but their precise form will not be important in what follows.
This is  the Lagrangian of a particle moving on a plane in the
presence of 
a constant magnetic field  of magnitude $N/2\pi $ perpendicular to the plane.
We also know that the 2-form $B$ fields are both compact
gauge fields because both the D-string and the F-string charge
is quantized.
Consequently, their zero modes
$ a$ and $ b$ are compact coordinates with period 
$ 2\pi$.

Now, without the coupling to the magnetic field, the Lagrangian
clearly has a $ U(1)$ shift symmetry $ a\rightarrow a+c$ for an
arbitrary periodic constant $ c$.  The coupling to the magnetic
field breaks this $ U(1)$ to a $ \bZ_N$ subgroup. This follows from the
 the fact that $ e^{-\int  L}$ would be invariant under the shift of $ a\rightarrow a +2\pi/N$ because $ b$ has period $ 2\pi$.  
It will
suffice for our purposes to consider the limit of large
magnetic field and drop the kinetic terms. The remaining
Lagrangian is automatically first order and we can readily identify
$b$ as the momentum conjugate to $ a$. They would 
satisfy the Heisenberg commutation relation
\beq
\label{comm2}
[{\hat a},{\hat b}] =2\pi i/N.
\eeq
Using this commutator it then follows that in our original problem we
have the operator relation 
\beq
\exp{(i k_3 \int_{\Sigma_{12}} {\hat B}^{R} )}
\exp{(i\int_{D_{3r}} {\hat B}^{NS})} =
\exp{(i k_3 \int_{\Sigma_{12}} {\hat B}^{R} )}
\exp{(i\int_{D_{3r}} {\hat B}^{NS})} \exp{({2\pi i k_3\over N})}
\eeq
which is identical to the relation eq.(\ref{comm}) that we were
seeking.
It implies the operator identification
\beq
\label{operator}
{\hat \Omega}[k_3] \leftrightarrow \exp{(i k_3 \int_{\Sigma_{12}} {\hat B}^{R} )}.
\eeq
A state in supergravity with  a well defined flux $ e_3$
is an eigenstate of this operator with 
\beq
\label{defel}
\int_{\Sigma_{12}} B^{R} = {2 \pi e_3 \over N}.
\eeq

By an analogous reasoning one can see that 
turning on $m_3$ units of magnetic flux in the $x_3$ direction then
corresponds 
to turning on an expectation value 
\beq
\label{defmag}
\int_{\Sigma_{12}} B^{NS}= {2\pi m_3 \over N}  \,.
\eeq
This also follows from the  $SL(2, \bZ)$ duality symmetry which in the
Type-IIB theory  interchanges $ B^{R}$  and $ B^{NS}$ and  in the
Yang-Mills theory on the D3-brane worldvolume
exchanges electric and magnetic fluxes.
Under this duality the Wilson loop $ \cA(C)$ that creates electric
flux tube gets interchanged with 
the  't Hooft loop ${\cal B} (C)$ which creates a magnetic flux tube.
In the supergravity theory, ${\cal B} (C)$ is thus 
a boundary of a D-string worldsheet.
In general, turning on the fluxes $ e_i$ and  $ m_i$ in the Yang-Mills theory
corresponds in the supergravity to turning on the expectation values
\beq
\label{general}
\int_{\Sigma_{ij}} B^{R} = {2 \pi \epsilon_{ijk}e_k \over N},\qquad
\int_{\Sigma_{ij}} B^{NS} = {2 \pi \epsilon_{ijk} m_k \over N}.
\eeq

\subsection{\subsecfont Finite Temperature Partition Function}

Having identified the discrete fluxes in the supergravity description 
we now turn to calculating the finite temperature partition function. 
When the 2-form gauge potentials are zero, we saw in section 2.1 that the 
relevant configuration in the supergravity theory is a black hole. 
What happens when the gauge potentials are turned on? 
A careful quantization of the zero modes 
as in \cite{AhWi} shows that, when the gauge potentials are constant
and take the quantized values,  there is no extra cost in energy. 
In turn this means that the 
metric, the 5-form field strength,  and the dilaton stay unchanged from 
their values in the black hole background.  
Thus the discrete fluxes do 
not change the classical behavior of the black hole and the related 
thermodynamic quantities. 

Once quantum fluctuations around the black holes background are included
though a dependence on the fluxes does arise. These quantum fluctuations
include Euclidean fundamental  string worldsheets that wrap around
non-trivial 2-cycles in the bulk geometry. These can be viewed as
F-string instantons and in this instanton sector the partition
function depends on the 
expectation value of the 2-form
gauge potential  $B^{NS}$. One can also have D-string instantons
that are sensitive to $ B^{R}$ and more generally $ (p, q)$-string
instantons \cite{Schw,Witt3}.
The discrete fluxes thus  manifest themselves in the dual 
supergravity theory as quantum hair \cite{BGHHS,CPW}.

Let us first consider the case where 
a single component of magnetic flux $m_{3}$ is turned on. 
{} From eq.(\ref{defmag}), we see that the quantum fluctuations of 
relevance then will involve fundamental
string Euclidean worldsheets that wrap around $ \Sigma_{12}$. 
We can organize
the partition function as a sum over configurations that
wrap $ n$ times around the 2-cycles:
\beq
\label{partfunc}
Z=Z_{sugra} + \sum_{n }Z_{(sugra + n \, strings )} 
\eeq
The leading term $Z_{sugra}$ is the contribution of the pure supergravity theory 
without any extra strings. It is determined by the black hole solution,
eq.(\ref{solbh}), to be  :
\beq
\label{z0}
Z_{sugra}=e^{(c_2 N^2 T^3 L^3)}, 
\eeq
where $c_2={\pi^2 \over 8}$. 
To precisely calculate the next term $Z_{(sugra + 1 string)}$ 
we would really need to know  Type-IIB string theory on
the black hole background or at least the sigma model
in the Green-Schwarz
formalism. We will estimate here the cost of including 
the string instanton using the Nambu-Goto action:
\beq
\label{z1}
Z_{(sugra + 1 string)}=e^{(c_2 N^2 T^3 L^3)} \int DX e^{-S_{NG} + i \int B^{NS}}.
\eeq
The dominant contribution that is  sensitive to the magnetic flux 
eq.(\ref{defmag}) is obtained by the  minimum area surface 
$ \Sigma_{12}$ that extremizes 
the Nambu-Goto action.
{}From eq.(\ref{solbh}) we see that this surface arises from a string world
sheet located at the horizon in the radial direction which spans 
the $x_1-x_2$ plane at some definite value of $x_3$.

The area of this minimal surface in units of the fundamental string tension is,
\beq
\label{areamin}
A_{H}/(2 \pi \alpha') = { 1 \over 2 \pi}~
                         {r_{H}^2 L^2 \over R^2 \alpha'}.
\eeq

Including the effects of an anti-string (or equivalently a string oppositely 
wound in the $x_1-x_2$ plane) gives:
\beq
\label{fullz1}
Z_{(sugra + 1 string)}=e^{(c_2 N^2 T^3 L^3)} ~ 2 C_F~ 
                          e^{-A_{H}/(2 \pi\alpha')}~
\cos({2 \pi m_{3} \over N}).
\eeq
The coefficient $C_F$ arises in the string path integral 
from fluctuations around the
minimal area surface.  It includes a contribution, $LT$, due to the zero mode
associated with the location of the surface in the $x_3$ direction, and the
contribution due to non-zero modes as well. We also need to include
the fluctuation of the fermionic modes.  In particular, we need to worry
if there are fermionic zero modes that might make the contribution vanish
as a consequence of spacetime supersymmetry.
We expect however  that 
the zero modes of the spacetime Green Schwarz fermions will be lifted
because the thermal boundary condition breaks supersymmetry  
completely.

Using eq.(\ref{temp}), eq.(\ref{areamin}), the minimum area 
can be re-expressed as:
\beq
\label{expsup}
e^{-A_{H}/( 2 \pi \alpha')}= e^{-\pi^{3/2} \sqrt{g_sN} T^2 L^2}.
\eeq
Not surprisingly the 
coefficient $\pi^{3/2} \sqrt{g_sN} T^2$ is  identical to the
string tension of the electric flux tube in the finite temperature 
 $N=4$ theory.  From the supergravity point of view,
the  theory has stable flux tubes
 because at finite temperature the classical geometry is given
 by the black hole geometry, eq(\ref{solbh}), in which the value of the
 metric parallel to the brane goes to a minimum non-zero value at the 
horizon \cite{Witt1, BISY}. 
Thus the behavior of  large Wilson loops is determined by the geometry
at the horizon  which also determines the exponential factor,
eq.(\ref{expsup}), above.

The contributions to the partition function from the single string instanton
 in the case of a general electric and magnetic
flux can now be written down by including the contributions of 
F and D strings wrapping the corresponding two-cycle:
\beqa
\label{gencase}
Z_{(sugra + 1 string)}[m_j, e_i]&=&e^{(c_2 N^2 T^3 L^3)}  [ 2 C_F
e^{-\pi^{3/2} \sqrt{g_sN} T^2 L^2} \sum_{j}\cos({2 \pi m_{j} \over N}) \nonumber \\
&& + 
2 C_D e^{-\pi^{3/2} \sqrt{N/g_s} T^2 L^2}  \sum_{i} \cos({2 \pi e_i \over N})].
\eeqa
There are subleading contributions to eq.(\ref{gencase}) which arise
from string-worldsheets that 
either wrap  multiply  or   wrap more 
than one two-cycle simultaneously. However, these contributions are 
exponentially suppressed compared to those included in eq.(\ref{gencase}) 
 as long as
\beq
\label{supcon}
\sqrt{N g_s}~ T^2~ L^2~ \gg 1,~~  \sqrt{N/ g_s}~ T^2~ L^2~ \gg 1.
\eeq
For small $g_s$ the first condition is more restrictive. 
Physically, eq.(\ref{supcon}) means that the torus has a size much 
bigger  than the electric and magnetic string tensions. 
When this is true the contributions to the partition function from 
multi-string sectors are also suppressed. 
In addition, there are subleading 
contributions from $(p, q)$ string worldsheets which we discuss shortly in 
eq.(\ref{pqcase}). Eq.(\ref{gencase}) thus gives the 
leading
dependence of the partition function on the discrete fluxes. 
It is the main result of this section.

Some comments are now in order.
First, it is worth noting that the minimum area surface which gave the 
dominant contribution to eq.(\ref{gencase}) above
has a simple interpretation in the gauge theory as well. It corresponds
to an electric or magnetic flux tube, which is small at first but grows 
with time 
to span the full 2-cycle. 
Second, we have neglected terms which are down in the $\alpha'$ 
and $g_s$ expansions here. Thus our result pertains to the gauge theory
at large N and strong 't Hooft coupling. 
Third, the result above does not apply to the zero temperature
case. In  this limit the condition, eq.(\ref{supcon}), is no longer
met and the contributions from multiply-wound and multi-string states
cannot be neglected.  

Finally, to understand the $SL(2, \bZ)$ transformation properties of 
eq.(\ref{gencase})
one needs to include the effects of general $(p,q) $ strings as well. 
Doing so gives a partition function: 
\beq
\label{pqcase}
Z[m_i,e_j](\tau) =\sum_{i,j}~ \sum_{p,q}~ C_{p,q}(\tau)~ \exp{(-A T_{p,q}(\tau))} ~
\cos[{2 \pi \over N}(-p  \cdot m   + q \cdot e) ].
\eeq
Here, $\tau= \chi + i e^{-\phi}$ is the axion-dilaton field and 
$C_{p,q}(\tau)$ is the determinant of 
small fluctuations which can in general depend on $\tau$.  
The string tension $T_{p,q}(\tau)$ of the $(p,q)$ string 
and the area  of the two-cycle $A= 2 \pi^{5/2} \alpha'\sqrt{N}T^2L^2$
  are both measured in
the Einstein metric.  Both
$( e_i, m_i)$ and   $(p_i ,q_j)$ transform as  vectors under the $SL(2, \bZ)$
transformation, and $\tau$ transforms as usual by fractional
linear transformation. 
Keeping this in mind we see that the last two terms of eq.(\ref{pqcase})
are invariant. Determining the full transformation properties of $Z[m_i,e_j]$ 
requires additional information about $C_{p,q}$ though which is beyond the scope of this paper.

\section{\secfont Non-supersymmetric Gauge Theory in  Three dimensions}
At energies smaller than the temperature, the four dimensional theory
reduces to a non-supersymmetric theory in three dimensions. 
The fermions and scalars of the $N=4$ theory acquire temperature dependent
masses leaving the pure glue degrees of freedom at low energies. 
The results of the previous section can be reinterpreted to 
tell us about the behavior of this theory. Before proceeding though
 it is important to emphasize that the resulting
theory is quite different 
from the usual definition of three-dimensional QCD. 
{}From the three-dimensional point of view
 the temperature acts like a cutoff.
We mentioned in the previous section that the confining scale
in the theory is of order $(g_sN)^{1/4} T$.
Thus, in the limit of strong 't Hooft coupling
considered here, the theory is already strongly coupled at
the cutoff scale.
Despite these differences it is interesting to ask how the
three-dimensional theory 
behaves as a function of the discrete fluxes.
The resulting answers are analogous to those
obtained in the strong coupling expansion of lattice gauge theory which,
despite differences with the continuum theory, are often illuminating.

We denote the two spatial directions of the three-dimensional theory as 
$x_1,x_2$ and the time direction as $x_3$. There is one integer
$m_{3}$ which characterizes the  magnetic flux in the 
$x_1-x_2$ plane. There are two electric fluxes in the 
$x_1, x_2$ directions respectively, characterized by the 
integers, $e_1, e_2$. We are interested in the partition function of this 
theory
when both the spatial directions and the time direction are taken to
be circles of length $L$, and periodic boundary conditions are
imposed in all three directions.
As was discussed in \S2 the dependence on the electric flux $e_i$
is related by a Fourier transform to the partition function obtained
by introducing twisted boundary conditions in the $x_1-x_3$ 
and $x_2-x_3$ directions.

In the supergravity theory the background geometry describing the theory
 is of course still 
described by the black hole, eq.(\ref{solbh}). However, in the three-dimensional
setting the coordinate axis
have a somewhat different interpretation.
$x_0$ can be regarded as the extra spatial direction along 
which supersymmetry breaking boundary 
conditions are imposed, while $x_3$ is the Euclidean continuation 
of the time direction. 

The leading dependence of the partition function on
the fluxes is therefore 
obtained by keeping the first term in eq.(\ref{gencase}) above:
\beqa
\label{zthree}
Z[m_{3}, n_1,n_2]&=&e^{(c_2 N^2 T^3 L^3)}  ~ 2 C_F~
e^{-\pi^{3/2} \sqrt{g_sN} T^2 L^2} \nonumber \\
&& [\cos({2 \pi m_{3} \over N}) + 
\cos({2 \pi n_1 \over N}) + \cos({2 \pi n_2 \over N})]. 
\eeqa
Turning on the twists in the $x_1-x_2$, $x_1-x_3$ and $x_2-x_3$ planes,
corresponds in the supergravity theory to turning on expectations values
for the $B^{NS}$ field along the appropriate two-cycle, eq.(\ref{defmag}). 

Note that in eq.(\ref{zthree}) there is no dependence on 
$ B^{R}$ because turning on an expectation value for 
$B^{R}$ field corresponds in the gauge theory 
to  twisted boundary conditions in a 
plane involving $x_0$, the extra compactified direction,
which has no intrinsic
significance from the point of view of the three-dimensional theory.

\section{\secfont Non-supersymmetric Gauge Theory in Four Dimensions}
We turn next to discussing a non-supersymmetric four-dimensional theory. 
In analogy with our discussion of the previous section we will start
with a five-dimensional theory realized by considering the world volume
theory of $D4$ branes and break supersymmetry by compactifying one 
of the brane directions on a circle with supersymmetry breaking boundary
conditions \cite{Witt1, GrOo}.  We compute the partition function of this theory
as a function of the various electric and magnetic fluxes.

The supergravity solution is given by  the
near-horizon  geometry of a configuration of $N$ near-extremal 
D4 branes \cite{IMSY}. The metric is 
\beq
\label{d4metric}
ds^2=({r^{3/2} \over R^{3/2}})~ (1- {r_H^3 \over r^3}) dx_0^2 + 
     ({R^{3/2} \over r^{3/2}}) ~ {dr^2 \over (1- {r_H^3 \over r^3})}~+~  {r^{3/2} \over R^{3/2}}
(dx_1^2+dx_2^2 +dx_3^2+dx_4^2)~+~ \sqrt{rR^3} d\Omega_4^2,
 \eeq
and the dilaton is
\beq
\exp\left( {\phi} \right) = g_s \left( {r\over R} \right)^{3\over 4},
\eeq
with $R^3=\pi g_s N (\alpha')^{3/2}$. 
{}From the point of view of the four-dimensional theory,
$x_0$ is the extra spatial direction along which supersymmetry breaking
boundary conditions are imposed, $x_1,x_2,x_3$ are the 
three spatial directions and $x_4$ is the Euclidean continuation of the time 
direction of the four-dimensional theory.

We are interested in the partition function of the gauge theory when the 
the three spatial and Euclidean time directions are 
 each compactified on a circle of length
 $L$.  There are six independent two cycles in this case and correspondingly
six integers, modulo N, which characterize the twisted boundary conditions that
can be introduced. Three of these directly correspond to the magnetic flux,
while the remaining three, which correspond to twisted boundary conditions 
involving $x_4$, can be related to the electric flux by a Fourier transform.

By reasoning similar to that in \S2 we find that turning on 
twisted boundary conditions along any of these six two-cycles corresponds 
to turning on an expectation value for $\int B^{NS}$, with the integral
being evaluated along the two-cycle:
\beq
\label{fluxd4}
{2 \pi n_{IJ} \over N}= \int_{\Sigma_{IJ}} B^{NS},
\eeq
where $I, J$ go from $ 1, \ldots, 4$.
The Supergravity Lagrangian contains a coupling of the form:
\beq
\label{sugrac4}
\delta L= \int B^{NS} \wedge dC^3  \wedge dC^3,
\eeq
where $C^3$ denotes the $3$ form RR field under which the $D4$ brane is
magnetically charged. This coupling is analogous to the Chern-Simons term in eq.(\ref{lagra}) 
and ensures that the $\int B^{NS}$ can take only $N$ distinct values. 

The leading dependence of the partition function on the discrete fluxes
can now be determined. It is
\beq
\label{gcase4}
Z[n_{IJ}]= e^{S_{BH}}~[1+ {\tilde C}~ e^{-\Lambda^2 L^2}~ \sum_{IJ}
\cos({2\pi n_{IJ} \over N})].
\eeq
The overall exponential factor arises from the free energy of the 
black hole geometry; an explicit calculation yields that $S_{BH} \sim
N^2 (g_s N \sqrt{\alpha'}T)T^4L^4$.    
The second term arises from the one fundamental string instanton sector, 
the corresponding exponential factor 
comes from the minimal area of the 
world sheet which sweeps out the appropriate two-cycle.
By considering large Wilson loops one can easily show that the theory
confines. Reasoning entirely analogous to that in section 2.3 shows that the 
constant, $\Lambda^2$, 
 is the string tension of the 
electric flux tube in the theory. Numerically,  
\beq
\label{deflambda}
\Lambda^2= {32 \over 27}~\pi^3 ~(g_s N \sqrt{\alpha'} T) T^2.
\eeq
Finally, ${\tilde C}$ is the contribution of the determinant for fluctuations
about the minimal area surface in the string path integral, it 
includes a factor proportional to $L^2 T^2$ which arises from zero modes
which correspond to moving the two-cycle in the two directions transverse to
it.  
The equation Eq.(\ref{gcase4}), is the analogue of eq.(\ref{zthree}) in the 
three-dimensional
case. The dependence on the electric fluxes can be obtained by doing 
a Fourier transform in the $n_{I4}$ variables. 

A few comments are now in order. 
First, as in the finite temperature four-dimensional theory, 
the answer we obtain here has a simple interpretation in the gauge theory.
The leading dependence on the discrete fluxes arises from electric flux
tubes whose world volume sweeps out  the relevant two-cycle of the torus. 
Second, eq.(\ref{gcase4}) is the leading contribution for a big torus,
i.e.,  when $\Lambda L \gg 1$. 
Subleading terms arise from multiply wrapped and multi string configurations. 
Third, we have neglected both $\alpha'$ corrections and $g_s$ 
corrections here. 
It is important to emphasize that neglecting the $\alpha'$ corrections
in particular requires us to work in a region of coupling constant
space where the four-dimensional  theory is much different from ordinary QCD.
Extra Kaluza-Klein states in the theory have a 
mass of order the temperature $T$. The 't Hooft  coupling of the four-
dimensional theory at the cutoff scale, $T$,  is 
$\lambda^2= g_s N \sqrt{\alpha'} T$. Keeping the curvature small 
in string units requires, 
 $\lambda^2 \gg 1$. Thus the four-dimensional
theory is already strongly coupled at the cutoff scale \footnote{
Alternatively, we see from eq.(\ref{deflambda}) that when $\lambda^2 \gg1$,
the string tension is bigger than the cutoff scale $T$.}. 
Finally, from the point of view of the five-dimensional theory there can be 
additional twists along planes involving the temperature direction. 
The dependence on such twists arises due to D-2 branes whose world volume
is a three cycle dual to the 
two-cycle along which the twisted boundary
conditions are turned on. These twists do not have any intrinsic
significance from the point of view of the four-dimensional theory.  

The 4-torus  has a geometric $SL(4, \bZ)$ duality symmetry and it is 
easy to verify that the answer obtained here, eq.(\ref{gcase4}), has that 
symmetry too once the discrete fluxes $n_{IJ}$ are appropriately transformed. 
The significance of this duality symmetry was first realized by 't Hooft who 
formulated the conditions for invariance under it in terms of various 
duality relations. Of particular interest  are  $90^{\circ}$ rotations in the Euclidean theory,
involving the time and one space directions, which exchange some of the electric
and magnetic fluxes.  The corresponding 't Hooft duality relations place strong 
constraints on the partition function which in turn constrain the phase structure of the 
gauge theory. It is worth sketching out how these general considerations apply in the present context. 
 Let $Z[\vec e, \vec m ]$ denote the 
partition function as a function of the electric and magnetic fluxes ( this is 
a discrete Fourier transform of eq.(\ref{gcase4})) and let us consider it in the the infinite volume limit,
$L \rightarrow \infty$. One can show that the duality relations allow for a solution
which in this limit has the behavior:    
$Z[0, \vec m]/Z[0,0] \rightarrow 1$, while 
$Z[\vec e , \vec m] /Z[0,0] \rightarrow 0$. 
This implies that the free energy of a purely magnetic flux tube goes to zero 
in this limit whereas that of an  electric flux tube diverges.
Therefore, only electric flux tubes are 
stable and  the theory is confining.  
In the large $N$ limit under discussion here, we know
beforehand from a direct computation of the Wilson loop that the theory confines. Even so, it is 
revealing to explicitly compute the limiting behavior of $Z[0, \vec m]/Z[0,0]$  and $Z[\vec e, \vec m] /Z[0,0]$.
 One finds that it is of the form
mentioned above, consistent with confinement.

We end with a brief discussion of another approximation to QCD,
 the strong coupling expansion in lattice gauge theory. The partition
function as a function of the discrete fluxes can be easily computed
in this approximation and goes as
 $Z[n_{IJ}] \sim e^{-\Lambda^2 A}
\, \cos({2 \pi n_{IJ} \over N})$,
where $\Lambda$ is the string tension. This is in agreement with eq.(\ref{gcase4}). 
Let us briefly review how this result is obtained. 
The fundamental degrees of freedom in lattice gauge theory are link variables. The action in terms of these
is defined by taking the product of all the links around a plaquette. The leading contribution to 
the partition function which depends on the twisted boundary conditions then arises
from tiling a two cycle of the torus minimally. This gives rise to the exponential dependence on the 
area.  Once the two cycle is tiled the integrals over the interior link variables can be done
simply using the fact 
that\footnote{Here
we restrict ourselves to the  large N limit, in general there is a finite $N$  correction to this.}
 $\int dU U^{i}_j (U^{\dagger})_k^l = \delta^i_k ~\delta^l_j$. Finally, the integrals over the edge link variables can be done after taking into account the twisted boundary
conditions. Summing over both orientations for tiling gives the cosine dependence.  
In fact, intuitively, one would expect any confining theory to give rise to a dependence on the 
twisted boundary conditions of the form, eq.(\ref{gcase4}). Such a dependence can only arise from 
a non-local operator, the Wilson loop. This operator costs exponentially in the area in the 
confining theory and gives rise to the cosine term simply because it measure the flux by Stokes theorem.
It is reassuring that our calculation above agrees with this expectation and
with  the strong coupling expansion. 

\vskip 0.4in
\leftline{\secfont Acknowledgements}

We  thank Bill Bardeen, Sumit Das,
Sunil Mukhi, Ashoke Sen, and Spenta Wadia for useful discussions.
A. D. would like to acknowledge the hospitality of 
the ICTP, the organizers of the Extended
Workshop in String Theory, 
the  Enrico Fermi Institute and the Particle Theory group at
the University of Chicago where some of this work was completed. 
S.~P.~T. would like to thank the Theoretical Physics group at Fermilab
for hospitality during his visit.

\nc{\np}[3]{ {\em Nucl.\ Phys. }{\bf #1} (19#2) #3}
\nc{\pl}[3]{ {\em Phys.\ Lett. }{\bf #1} (19#2) #3}
\nc{\pr}[3]{ {\em Phys.\ Rev. }{\bf #1} (19#2) #3}
\nc{\prep}[3]{ {\em Phys.\ Rep. }{\bf #1} (19#2) #3}
\nc{\prl}[3]{ {\em Phys.\ Rev.\ Lett. }{\bf #1} (19#2) #3}

\end{document}